\documentclass[a4paper,conference,10pt]{IEEEtran}
\IEEEoverridecommandlockouts
\usepackage{amsmath,amssymb,amsfonts}
\usepackage{graphicx}
\usepackage{balance}
\usepackage{geometry}
\def\BibTeX{{\rm B\kern-.05em{\sc i\kern-.025em b}\kern-.08em
    T\kern-.1667em\lower.7ex\hbox{E}\kern-.125emX}}

\usepackage{xcolor}

\renewcommand{\t}{^\top}
\renewcommand{\matrix}[1]{\begin{bmatrix}#1\end{bmatrix}}
\newcommand{\norm}[1]{\left\lVert#1\right\rVert}

\usepackage{tikz}

\newcommand\acceptedtext{%
  \footnotesize © 2023 IEEE.  Personal use of this material is permitted.  Permission from IEEE must be obtained for all other uses, in any current or future media, including reprinting/republishing this material for advertising or promotional purposes, creating new collective works, for resale or redistribution to servers or lists, or reuse of any copyrighted component of this work in other works.}

\newcommand\acceptednotice{%
\begin{tikzpicture}[remember picture,overlay]
\node[anchor=south,yshift=10pt] at (current page.south) {\fbox{\parbox{\dimexpr0.65\textwidth-\fboxsep-\fboxrule\relax}{\acceptedtext}}};
\end{tikzpicture}%
}

\begin{document}
\title{Artificial intelligence-based predictive fuel blending control for flare gas mitigation}

\author{\IEEEauthorblockN{J. Kir Hromatko\IEEEauthorrefmark{1},  Š. Ileš\IEEEauthorrefmark{1}, R. Huljev\IEEEauthorrefmark{2}, V. Vučković\IEEEauthorrefmark{3}} \
\IEEEauthorblockA{\IEEEauthorrefmark{1} University of Zagreb, Faculty of Electrical Engineering and Computing, Zagreb, Croatia \\
E-mail: \{josip.kir.hromatko, sandor.iles\}@fer.hr}

\IEEEauthorblockA{\IEEEauthorrefmark{2} Genuyn, Office 9, Dalton House, 60 Windsor Avenue, London, United Kingdom, SW19 2RR\\
E-mail: rube@genuyn.com}

\IEEEauthorblockA{\IEEEauthorrefmark{3} Popravak Brodskih Motora d.o.o.,
Svilno bb, 51219 Čavle (Rijeka), Croatia \\
E-mail: velibor.vuckovic@pbm.hr}

\thanks{This work has been supported in part by the European Regional Development Fund under the project  KK.01.2.1.02.107.}}

\maketitle
\acceptednotice

\begin{abstract}
This paper describes a fuel blending algorithm based on artificial intelligence and model predictive control. A gas-fired power plant was modeled using physical laws and on-site measurements. A neural network is used to calculate the methane number of the fuel and determine the fuel blending ratio limits so that the methane number is within the limits specified by the engine manufacturer. A model predictive controller adjusts the final blending ratio to meet safety requirements and minimize operating costs. The algorithm was tested in simulations with different scenarios and a reduction in both the operating costs and amount of flaring was observed.
\end{abstract}

\renewcommand\IEEEkeywordsname{Keywords}
\begin{IEEEkeywords}
fuel blending, methane number, neural network, model predictive control, flare mitigation
\end{IEEEkeywords}

\section{Introduction}

Flares are important safety devices that can burn off unwanted
gases due to their volatile nature concerning the amount, availability, and quality. However, if operated improperly, flaring can emit vast amounts
of pollutants, and pose significant environmental and health risks. It is highly desirable to achieve zero flaring via flare minimization and flare gas recovery. 
By integrating the flared gases into a fuel gas network, it is possible to reduce emissions, as well as conserve energy in refineries \cite{tahouni2016integration}.
A comparison of economic profitability for various scenarios for using flare gases is investigated in \cite{nezhadfard2020power}. According to the paper, the flare gases with higher flow rates and better gas composition in terms of the amount of hydrogen and hydrogen sulfide highly influence the profitability of using such gases in various applications.

In this paper, we investigate the possibility of using model predictive control and neural networks to use flare gases in a dual-fuel engine. 

Model predictive control (MPC) is a control strategy that uses a model of a system to predict its response to a given control input. An MPC algorithm repeatedly solves an optimization problem and computes the optimal control action that minimises the cost function and controls the states of the system to the desired operating point while satisfying the control input and state constraints. MPC is widely used in industrial and process control applications, where it is valued for its ability to handle constraints. For example, MPC is used to control nonlinear systems with constraints, such as internal combustion engines where multiple objectives must be satisfied \cite{norouzi2021model}, or in applications such as fuel blending of six different feed gases where the objective is to achieve fuel blends with specific limits on the different composition properties \cite{muller2011modelling}. 

Instead of industrial fuel blending, in this paper we investigate the possibility of blending the flaring gas with natural gas in order to use it with a dual-fuel engine (or a gas-powered generator - genset). However, one of the main quality requirements of natural gas as an engine fuel is the methane number (MN). This parameter indicates the fuel’s capability to avoid knocking (spontaneous combustion) in the engine. A higher MN value indicates a better fuel quality for gas engines. The required minimum MN value usually depends on the engine manufacturer. More and more engine manufacturers or require fuel with a minimum methane number of 80, while some still maintain methane number limits below 80 (most often 70) \cite{kuczynski2020impact}. In \cite{kreutzer2011evaluation}, a fuel blending system was used to vary the MN. With smaller MNs,  NOx and CO emissions increased while combustion stability remained unaffected. 
 
The use of artificial neural networks for methane number prediction is reported in \cite{gupta2021predicting}, where the model was trained using 1202 different gaseous fuel compositions. The neural network model was compared to the AVL (Anstalt für Verbrennungskraftmaschinen List) model and it was reported that the model was able to predict the MN accurately (R = 0.999).

In this paper, the artificial neural network is used to predict the MN of different fuel blends based on the tail gas and natural gas with assumed known gas composition. The results from the neural network are used to find the limits on the fuel blending ratio which achieve the MN according to the motor specification. The model predictive control algorithm is used to minimize the price of the fuel blend and the amount of flaring while respecting several constraints on the variables of interest.

\section{Methods}
\subsection{Plant description}
The industrial process of interest in this work is a gas-fired power plant. The main component of the power plant is a generator set which runs on gas and produces approximately 1 MW of electrical power. Two gases are available for fueling the generator: natural gas, which is in constant supply and has to be paid for, and permeate gas, whose supply varies, but is available for free. The generator runs on a mixture of the two fuel gases and the mixing ratio is set manually. The permeate gas is usually of lower quality (i.e., contains less methane), which can affect the generation of power and should be considered when setting the mixing ratio. 

Additionally, a gas tank was installed on the permeate gas stream to act as a buffer and mitigate the effect of variations in availability. Since the pressure in the tank needs to be limited, a safety mechanism which flares excess gas in case of high tank pressure was implemented.

\begin{figure}
    \centering
    \includegraphics[width=.9\columnwidth,height=2.5cm]{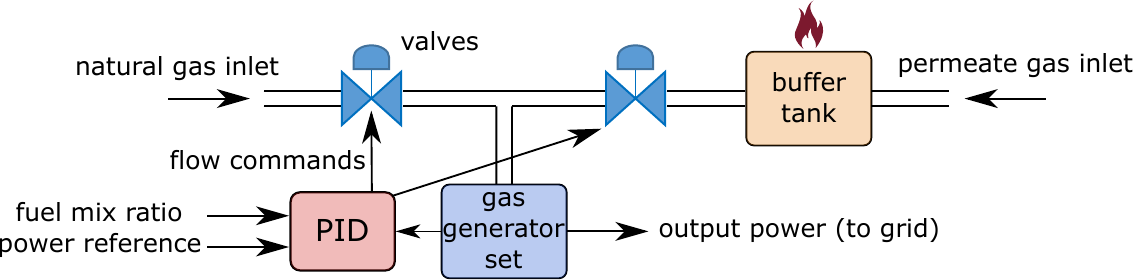}
    \caption{Illustration of the plant operation.}
    \label{fig:postrojenje}
\end{figure}

A simplified schematic of the power plant is given in Figure \ref{fig:postrojenje}.

\subsection{Plant model}
The power output of a fuel engine that runs on methane depends on the flow rate of the methane, as well as the engine's design and operating conditions. Generally, increasing the flow rate of methane to the engine will result in an increase in power output. However, there is a limit to how much methane can be safely and efficiently combusted within the engine, and at high flow rates, the engine may not be able to utilize the methane, resulting in decreased power output. Additionally, the engine's compression ratio, ignition timing, and other factors can also affect the engine's power output.

\begin{figure}
    \centering
    \includegraphics[width=.9\columnwidth]{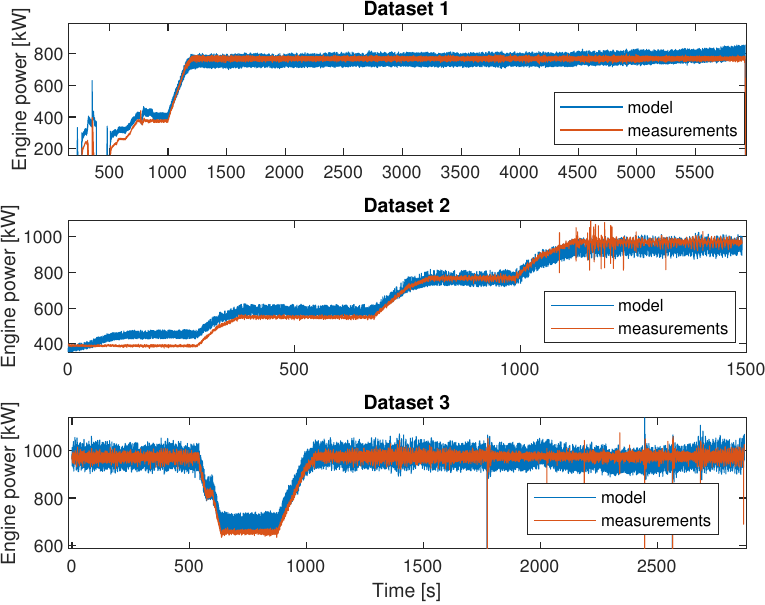}
    \caption{Recorded data and model fit.}
    \label{fig:power_fit}
\end{figure}

The focus of the work presented here was on controlling two variables of interest, the power output of the gas generator (engine) and the pressure in the gas tank (buffer). The actual power plant contains several control loops which ensure stable operation and allow for certain simplifications in control design. 

In modelling the gas generator, it was assumed that the output power is proportional to the amount of methane in the fuel, with a certain response time. The resulting model can be written as a differential equation:
\begin{equation}
    \frac{dP}{dt}=\frac{1}{\tau}\left(k_nq_n+k_pq_p-P\right)
    \label{eq:power_dot}
\end{equation}
where $P$ is the generated power, $q_i$ the mass flow of fuel component $i$ and $\tau$ the time constant of the power output. The notation $(\cdot)_n$ and $(\cdot)_p$ refers to natural and permeate gas quantities. The gains from fuel components to generated power, $k_n$ and $k_p$, generally depend on the fuel characteristics and the operating point (engine speed and torque), as well as other factors such as the engine efficiency. However, the on-site measurements enabled calculating the gains by a least squares fit and approximating them as constants. Figure \ref{fig:power_fit} shows the measurements and the power output calculations using the obtained gains, which were in the order of $10\pm 1\ kW/(L/s)$ for different datasets.

The tank pressure is affected by the incoming and outgoing permeate gas. The conservation of mass principle states that \cite{terma}: 
\begin{equation}
    \frac{dM}{dt}=\dot{m}_{in}-\dot{m}_{out}
\end{equation}
where $M$ is the total mass of a substance in a tank and $\dot{m}_i$ are the incoming and outgoing mass flows. Following the procedure in \cite{tank}, the derivative of the tank pressure can be expressed as:
\begin{equation}
    \frac{dp}{dt}=\frac{p}{\rho V}(q_i-q_p)
    \label{eq:pressure_dot}
\end{equation}
where $p$, $\rho$ and $V$ denote the gas pressure, density and volume, $q_i$ the incoming mass flow and $q_p$ the outgoing mass flow of permeate gas.

The available measurements indicate that the gas density is approximately constant and equal to 1 $kg/m^3$ for both gases so volumetric and mass flow can be used for calculations interchangeably.

\subsection{Neural network-based ratio limit calculation}
Gas engine manufacturers usually specify the desired limits for the methane number of fuel. Several methane number calculators are available as web \cite{wartsila_MN_calc} or mobile phone \cite{cat_MN_calc} applications, but using them in a control system requires a hardware implementation. The European Association of Internal Combustion Engine and Alternative Powertrain Manufacturers provides an open-source methane number calculator developed using Microsoft Excel and BASIC \cite{gitMN}.

\begin{table*}[!ht]
    \centering
    \caption{Permeate gas contents sampled during one year}
    \resizebox{\textwidth}{!}{
    \begin{tabular}{|c|cccccccccccc|}
    \hline
        component & 8.1.2021. & 3.2.2021. & 3.3.2021. & 6.4.2021. & 5.5.2021. & 7.6.2021. & 6.7.2021. & 10.8.2021. & 2.9.2021. & 4.10.2021. & 3.11.2021. & mean value \\ \hline
        C1 [mol\%] & 60.58 & 62.02 & 75.08 & 80.13 & 66.71 & 67.86 & 61.71 & 67.86 & 76.1 & 55.866 & 68.471 & 67.49 \\ 
        C2 [mol\%] & 1.28 & 1.44 & 1.63 & 4.21 & 1.26 & 1.36 & 8.23 & 2.07 & 6.53 & 10.159 & 6.751 & 4.08 \\ 
        C3 [mol\%] & 0.32 & 0.13 & 0.15 & 0.51 & 0.15 & 0.17 & 2.31 & 0.24 & 0.96 & 1.715 & 0.972 & 0.69 \\ 
        i-C4 [mol\%] & 0.1 & 0 & 0.05 & 0.12 & 0.05 & 0.05 & 0.45 & 0.04 & 0.17 & 0.294 & 0.161 & 0.14 \\ 
        n-C4 [mol\%] & 0.28 & 0 & 0 & 0.04 & 0.02 & 0.09 & 0.59 & 0.02 & 0.11 & 0.22 & 0.085 & 0.13 \\ 
        i-C5 [mol\%] & 0.16 & 0 & 0 & 0 & 0 & 0.04 & 0.18 & 0 & 0.05 & 0.082 & 0.033 & 0.05 \\ 
        n-C5 [mol\%] & 0.24 & 0 & 0 & 0 & 0 & 0.06 & 0.13 & 0 & 0 & 0.07 & 0 & 0.05 \\ 
        C6+ [mol\%] & 0.72 & 0.07 & 0.12 & 0.31 & 0.03 & 0.59 & 0.31 & 0.05 & 0.23 & 0.305 & 0.06 & 0.25 \\ 
        N2 [mol\%] & 4.12 & 4.58 & 6.1 & 6.48 & 7.45 & 7.65 & 1.58 & 7.66 & 4.77 & 2.312 & 4.077 & 5.16 \\ 
        CO2 [mol\%] & 32.21 & 31.77 & 16.87 & 8.2 & 24.33 & 22.14 & 24.51 & 22.08 & 11.09 & 28.977 & 19.39 & 21.96 \\ \hline
    \end{tabular}
    }
    \label{tab:gas_contents}
\end{table*}

\begin{figure}
    \centering
    \includegraphics[width=.95\columnwidth]{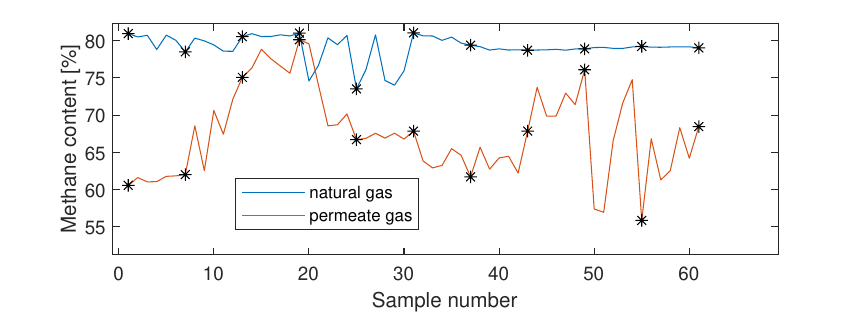}
    \caption{Gas methane contents dataset (stars indicate the measurements).}
    \label{fig:gas_contents_dataset}
\end{figure}

Since the methane number is a nonlinear function of the gas components and running on an embedded system is required, a neural network was trained for calculating the methane number. For this purpose, on-site gas contents measurements were used. The permeate gas measurements obtained during one year are given in Table \ref{tab:gas_contents}. The natural gas measurements from the same period are approximately constant and similar to those of permeate gas in April 2021. With this data, an augmented dataset of 110 samples was created with random linear combinations of adjacent samples, as shown in Figure \ref{fig:gas_contents_dataset}.

\begin{figure}
    \centering
    \includegraphics[width=.95\columnwidth]{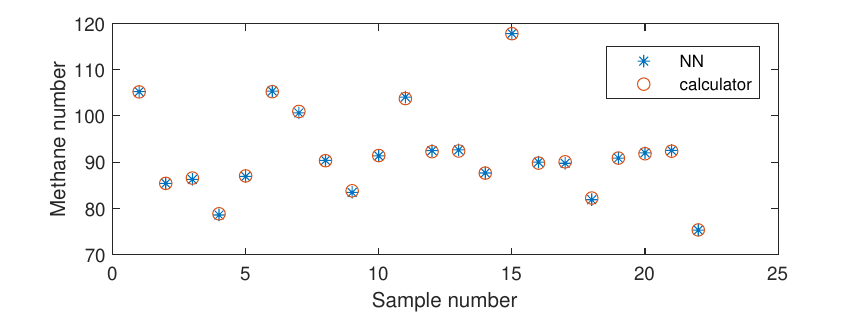}
    \caption{Neural network performance on the test set.}
    \label{fig:NN_performance}
\end{figure}

A relatively small network, with one hidden layer and two neurons with a radial basis activation function, was trained using Bayesian regularization and an 80/20 train/test split. It proved to be complex enough and performed well on the test set (MSE=0.0547), as shown in Figure \ref{fig:NN_performance}. The main difference between this network and the one presented in \cite{gupta2021predicting} is the network size; in this case, a much smaller network is sufficient since there is less variation in the dataset.

Given the contents of the natural and permeate gases, the neural network calculates the methane number for each one. The methane number of the engine fuel, however, depends on the mixing ratio of the two components. For simplicity, it was assumed that the contents of the mix can be calculated as:
\begin{equation}
    c_{mix}=rc_p+(1-r)c_n
\end{equation}
where $c_i$ are the contents of component $i$ and $r$ is the mixing ratio. The methane number of the mix is then calculated for different values of the mixing ratio and the boundary ratio values which satisfy the methane number limits (set by the manufacturer or the operator) are found.

Since the methane number is a nonlinear function of the components, it is possible that there are several mix ratio intervals which satisfy the methane number constraint. However, only one interval can be included in the optimal control problem formulation presented in the next section. In that case, the interval which contains the current mix ratio is chosen to avoid oscillations in plant operation. If the current mix ratio is not in any of the allowed intervals, the closest permitted value is chosen. Additionally, if the gas contents are such that the methane number constraint cannot be satisfied for any mix ratio, the ratio which will produce the best possible methane number (although outside the specifications) is selected. 

\begin{figure}
    \centering
    \includegraphics[width=.9\columnwidth]{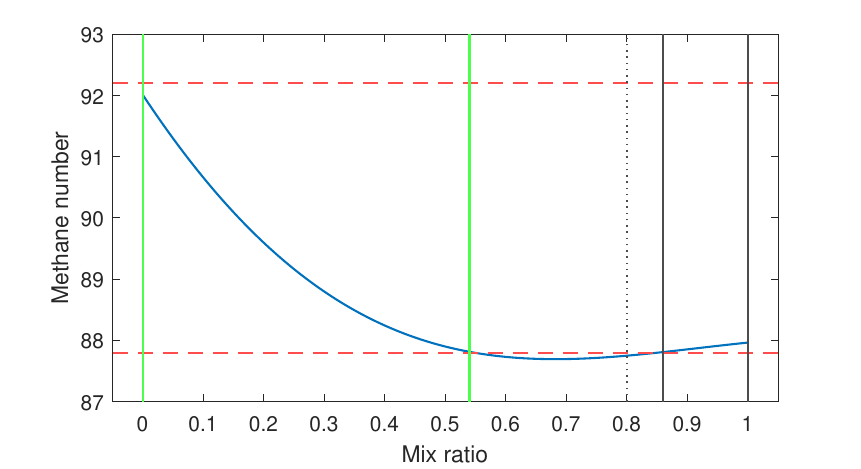}
    \caption{Ratio limit selection.}
    \label{fig:ratio_limits_interval}
\end{figure}

An illustration is given in Figure \ref{fig:ratio_limits_interval}, where the green line indicates the permitted mix ratio intervals, the dotted black line shows the current mix ratio and the full black lines indicate the selected mix ratio limits.

\subsection{Model predictive fuel mix ratio controller}
In order to minimize the use of gas flares and the operating costs, while limiting the tank pressure and power tracking error, a model predictive controller was used to determine the optimal fuel mix ratio. The reason for choosing the mix ratio instead of the individual gas flows as the output variable was the already existing engine controller, which determines the final flows according to the desired mix ratio and generated power.

Generally, a model predictive controller seeks the control sequence such that the user-defined cost function (which contains, e.g., energy usage, reference tracking error, safety limit violations) is minimized. The total cost function is calculated over a prediction horizon of open-loop behavior, consisting of several discrete time steps. Also, the choice of optimal control values is subject to state and control constraints (e.g., safety boundaries and actuator limitations).

\subsubsection{Discrete-time model}
Firstly, a discrete-time model of the plant is needed for MPC. Expressions \eqref{eq:power_dot} and \eqref{eq:pressure_dot} lead to the following state-space representation:
\begin{equation}
    \matrix{P\\p}^+=\matrix{1-\frac{T_s}{\tau} & 0\\0 & 1}\matrix{P\\p} + \matrix{\frac{T_s}{\tau}k_n & \frac{T_s}{\tau}k_p \\ 0 & -\frac{T_sp}{\rho V}} \matrix{q_n\\q_p}
    \label{eq:ss}
\end{equation}
where $T_s$ denotes the chosen sampling time. Note that the incoming flow of permeate gas, $q_i$, is not included in this model since it is not measured/controlled and can be treated as an unknown disturbance. Also, it is assumed that both the generator power and the tank pressure are measured. Finally, the required output, optimal mixing ratio, can be obtained from the optimal gas flows, $q_n$ and $q_p$, as $r=q_p/q_n$. The resulting model \eqref{eq:ss} is a discrete linear time-varying system:
\begin{equation}
    x^+=Ax+B(x)u,\quad x=[P\ p]\t,\quad u=[q_n\ q_p]\t
\end{equation}
which can be used in the MPC design.

\subsubsection{The objective function}
The objective function indicates which variables are important, i.e., should be minimized in the optimization. In this instance, the primary aim of the controller is to minimize deviations from the reference and the operating cost. The objective function can then be formulated as:
\begin{equation}
    V(x,u,r) = \sum_{i=0}^N \norm{x_i-r}^2_Q + \norm{u_i}^2_R
    \label{eq:mpc_cost}
\end{equation}
where $N$ is the prediction horizon, $x=[x_0,\ldots,x_N]\t$ and $u=[u_0,\ldots,u_N]\t$ the state and control sequences, $r$ the state reference and $Q$ and $R$ positive semi-definite weight matrices. The notation $\norm{v}_M^2$ denotes the quadratic form $v\t Mv$. 

Reference tracking errors of engine power and tank pressure can be penalized individually by choosing $Q$ to be diagonal. Also, operating cost minimization can be achieved by setting:
\begin{equation}
    R=\matrix{r_n & 0 \\ 0 & r_p},\quad r_n\gg r_p
\end{equation}
which will also result in mix ratio maximization, i.e., permeate gas will be used as much as possible and flaring will be minimized.

\subsubsection{Constraints}
Ideally, the optimal control problem should be formulated as a linear or quadratic program, since there are efficient methods and solvers for solving them \cite{boyd_vandenberghe_2004}. To achieve this, the control input matrix in \eqref{eq:ss} is assumed to be constant during the horizon, using the current tank pressure measurement. If the tank pressure does not change significantly during the prediction horizon, this simplification should not affect the results. Another option would be to use the controller predictions from the previous time step. The system dynamics can then be included as linear constraints in the optimization problem.

Additionally, the gas flows $q_n$ and $q_p$ should be constrained between zero and maximum flows, $q_{n,max}$ and $q_{p,max}$. The mix ratio constraints from the neural network can be added with the following expression:
\begin{equation}
    r_{min} \leq \frac{q_p}{q_n} \leq r_{max}
\end{equation}
Since the natural gas flow is nonnegative, this can be rewritten as a linear inequality:
\begin{equation}
    q_nr_{min} \leq q_p \leq q_nr_{max}
\end{equation}

If the optimization argument includes the state, an initial state constraint is needed such that the predictions start from the current measurement. Finally, the engine power and tank pressure can also be constrained to lie between the minimum and maximum allowed values, $x_{min}\leq x\leq x_{max}$.

\subsubsection{The optimization problem}
The resulting optimization problem can be formulated as:
\begin{subequations}
\begin{align}
    \underset{x,u}{\text{min}}&\quad V(x,u,r)=\sum_{i=0}^N \norm{x_i-x_r}^2_Q+\norm{u_i}^2_R\\
    \text{s.t.}&\quad x_i^+=Ax_i+Bu_i\\
    &\quad x(0)=x_0\\
    &\quad x_{min} \leq x_i \leq x_{max}\\
    &\quad u_{min} \leq u_i \leq u_{max}\\
    &\quad u_{1,i}r_{min} \leq u_{2,i} \leq u_{1,i}r_{max},\quad\forall i\in[1,N]
\end{align}
\end{subequations}
The cost function is quadratic in the optimization variables and all constraints are linear. Therefore, the problem can be written as a standard quadratic program:
\begin{subequations}
\begin{align}
    \underset{x}{\text{min}}&\quad \frac{1}{2}x\t Px+q\t x\\
    \text{s.t.}&\quad l \leq Ax \leq u
\end{align}
\end{subequations}
and the optimal solution can be obtained using a QP solver such as OSQP \cite{osqp} or MATLAB's quadprog \cite{quadprog}.

\subsection{The simulation model}
A simulation of the plant and the control algorithm was developed using MATLAB and Simulink. In the simulation, gas contents are periodically sampled from the dataset used to train and test the neural network, i.e., the on-site measurements. The neural network uses these measurements to calculate the allowed fuel mix ratio. The mix ratio limits are given to the MPC, which outputs the optimal mix ratio that satisfies the imposed constraints and minimizes the specified cost. Finally, this ratio is used by a lower-level PID controller that sets the individual flows of the gases in order to achieve the desired engine power output. 

The control loop runs with a sampling time of 1 second, while the engine power and tank pressure dynamics are simulated using continuous-time models \eqref{eq:power_dot} and \eqref{eq:pressure_dot}. An upper bound on the tank pressure is not imposed since the excess gas can be flared. Also, the incoming permeate gas is modeled as a uniformly distributed noise of moderate amplitude. 

\begin{figure}
    \centering
    \includegraphics[width=.9\columnwidth]{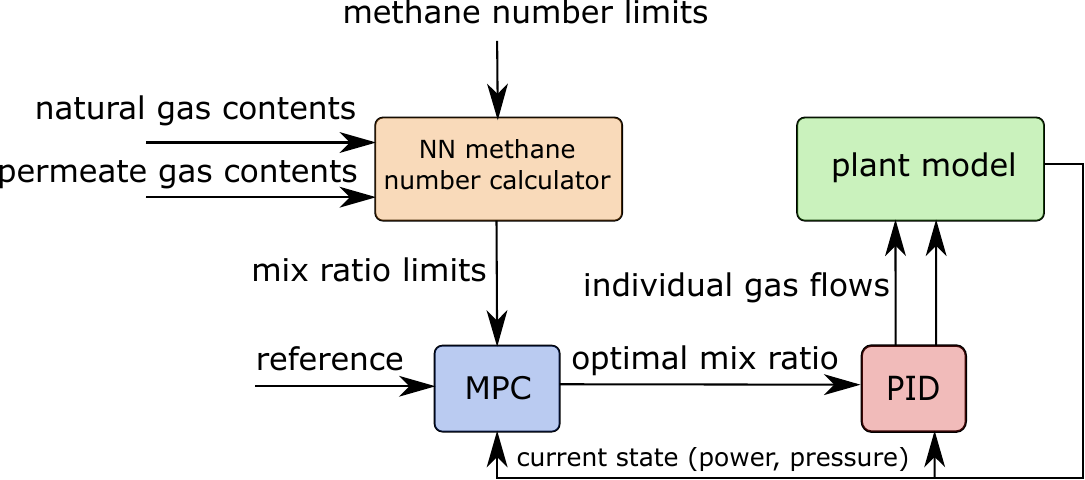}
    \caption{The simulation scheme.}
    \label{fig:simulacija}
\end{figure}

The simulation scheme is shown in Figure \ref{fig:simulacija} and the most important parameters are given in Table \ref{tab:sim_params}.

\begin{table}
    \centering
    \caption{Simulation parameters}
    \resizebox{\columnwidth}{!}{
    \begin{tabular}{c c c c }
         \hline
         Parameter & Value & Unit & Description\\
         \hline
          $q_{max}$ & $ 300 $ & $L/s$ & maximum fuel flow\\
          $q_{i,max}$ & $ 0.1667 $ & $L/s$ & maximum incoming permeate gas flow\\          
          $\tau$ & $ 20 $ & $s$ & power response time constant\\
          $V$ & $ 35 $ & $m^3$ & buffer tank volume\\
          $N$ & $ 20 $ & $-$ & MPC prediction horizon\\
          $Q_P$ & $1/1000^2$ & $1/(kW)^2$ & MPC power tracking error weight\\
          $Q_p$ & $1/60^2$ & $1/(kPa)^2$ & MPC pressure tracking error weight\\
          $R_n$ & $1\cdot10^{-4}$ & $1/(L/s)^2$ & MPC natural gas flow cost\\
          $R_p$ & $1\cdot10^{-6}$ & $1/(L/s)^2$ & MPC permeate gas flow cost\\
          $T_s$ & $1$ & $s$ & control loop sampling time\\
          \hline
    \end{tabular}}
    \label{tab:sim_params}
\end{table}

\section{Results}
The primary objective of the described control scheme was to automatically set the mixing ratio of the two gases. Therefore, closed-loop simulations were compared with the case where the mix ratio is set manually and constant. For calculating the cost of fuel, it was assumed that the unit price of natural gas is 1 and that of the permeate gas is 0.3. Two different scenarios were considered. 

In the first scenario, the natural and permeate gas contents were obtained by blending the points obtained from on-site measurements (Figure \ref{fig:gas_contents_dataset}). In this case, the natural gas content varies slowly and the methane content in the natural gas is approximately constant, while the permeate gas is more volatile.

In the second scenario, the gas contents are randomly selected from the combined dataset, and this scenario represents a mixing of two volatile gasses where it is more difficult to meet the methane number limits.

\begin{figure}
    \centering
    \includegraphics[width=.95\columnwidth]{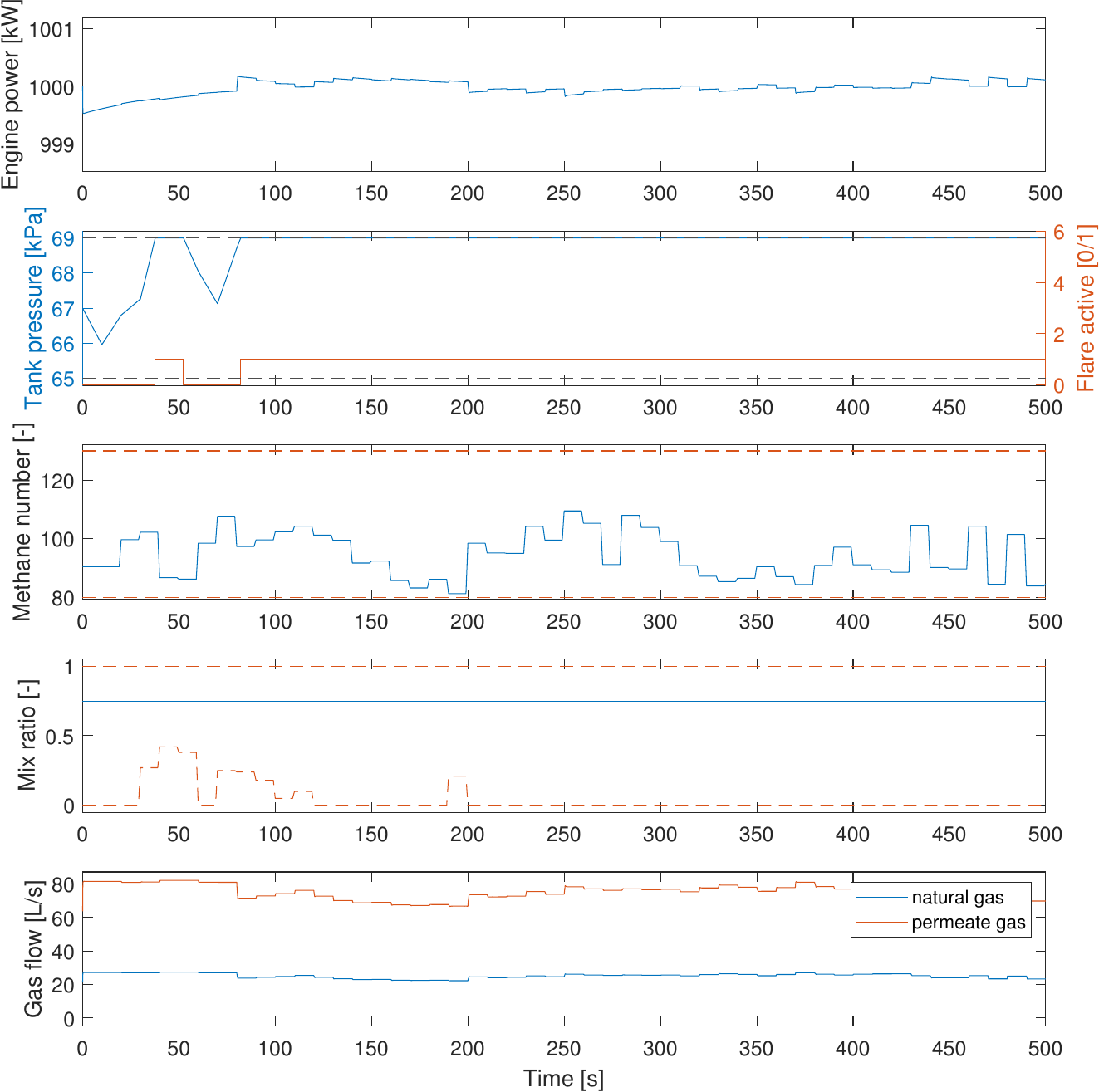}
    \caption{Simulation results with a manually set mix ratio and sequentially chosen gas contents.}
    \label{fig:sim2_manual}
\end{figure}

\begin{figure}
    \centering
    \includegraphics[width=.95\columnwidth]{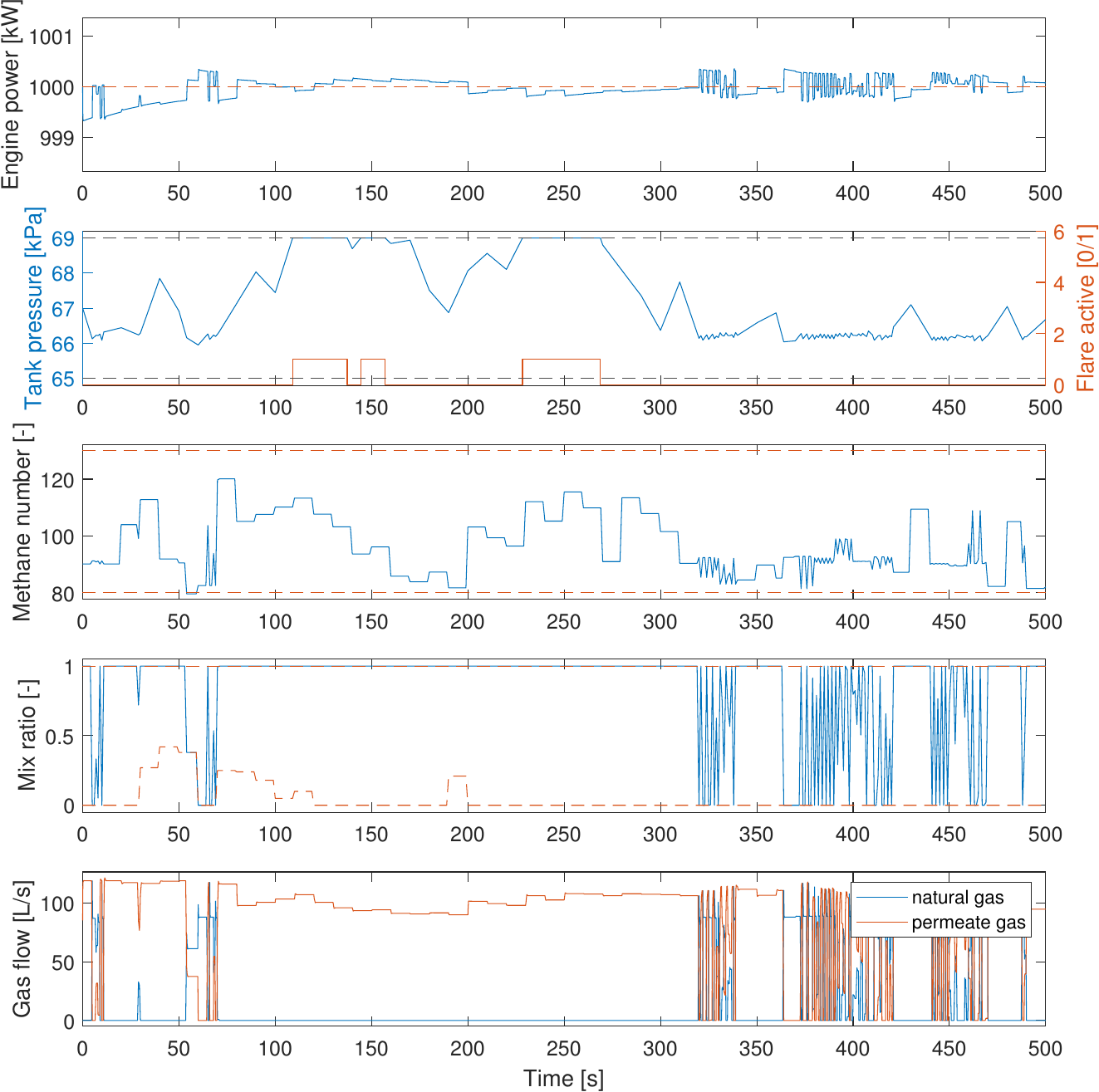}
    \caption{Simulation results with an automatically set mix ratio and sequentially chosen gas contents.}
    \label{fig:sim2_control}
\end{figure}

Figures \ref{fig:sim2_manual} and \ref{fig:sim2_control} show the simulation results for the first scenario with the gas contents dataset shown in Figure \ref{fig:gas_contents_dataset}. In both cases, engine power, tank pressure and methane number are within the specified limits. However, in the second case, the control input oscillates more, while the total fuel cost is reduced by 16\% and the flaring time is reduced by 81\%.

\begin{figure}
    \centering
    \includegraphics[width=.95\columnwidth]{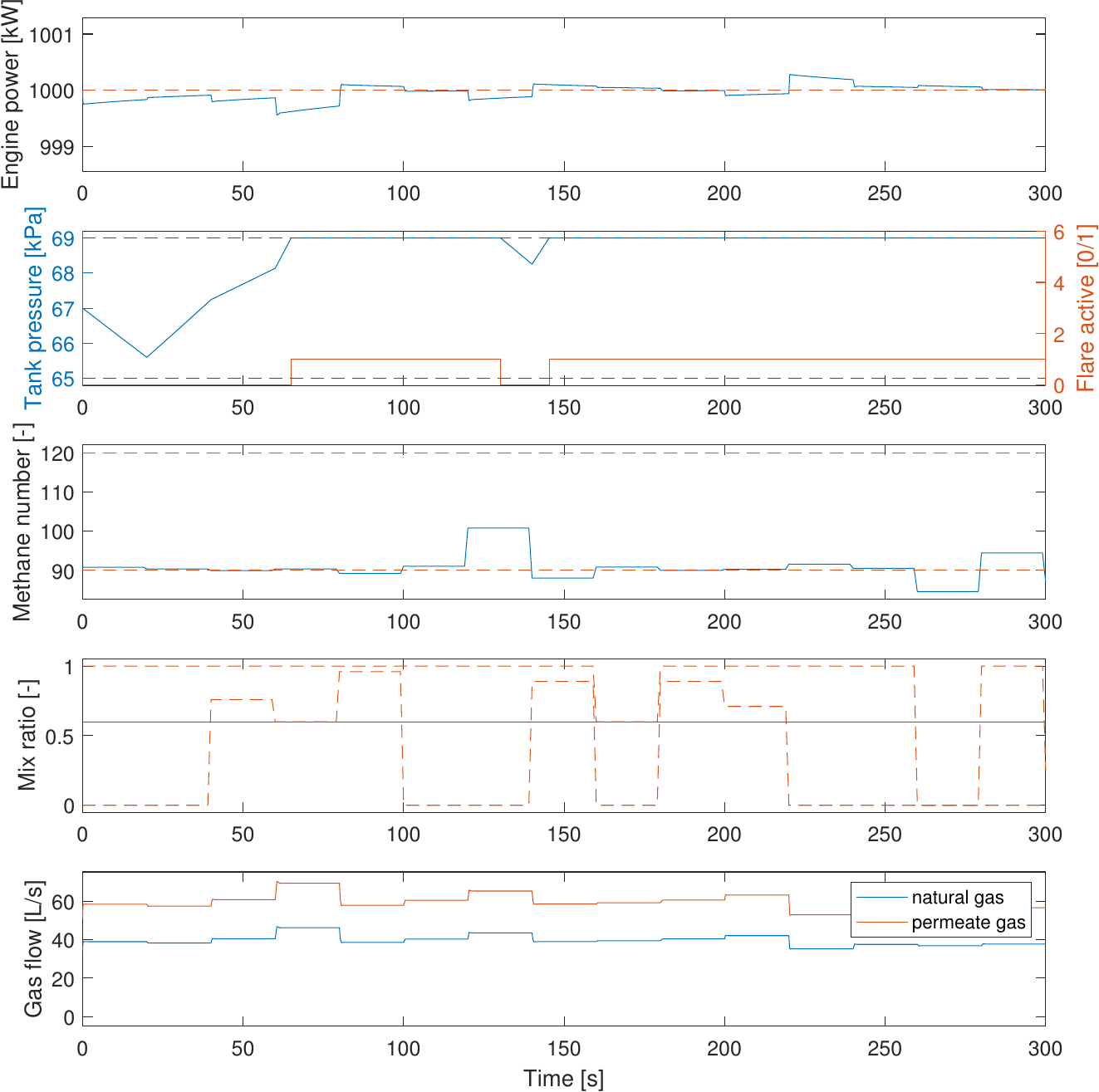}
    \caption{Simulation results with a manually set mix ratio and randomly chosen gas contents.}
    \label{fig:sim_manual}
\end{figure}

\begin{figure}
    \centering
    \includegraphics[width=.95\columnwidth]{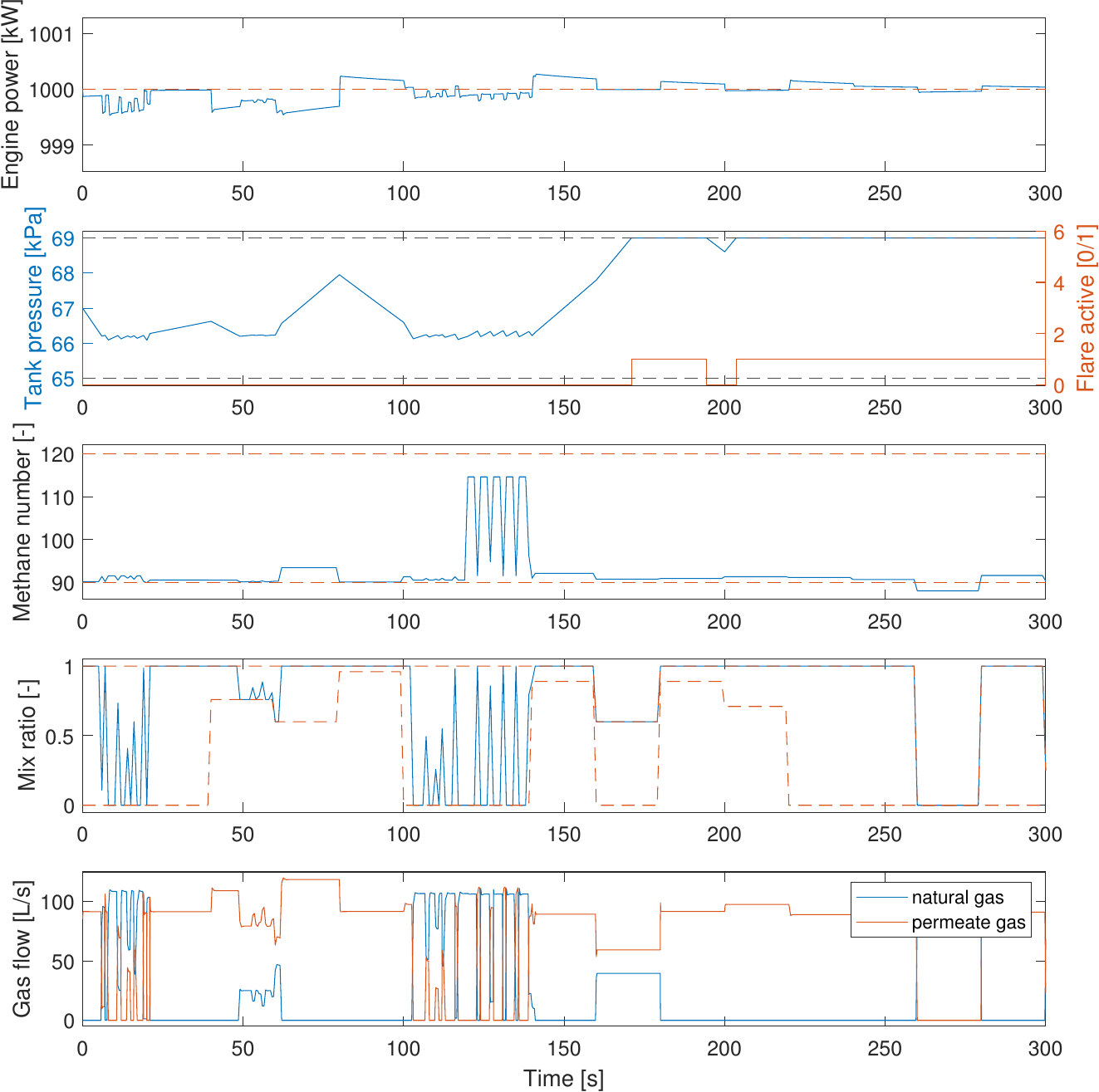}
    \caption{Simulation results with an automatically set mix ratio and randomly chosen gas contents.}
    \label{fig:sim_control}
\end{figure}

Figures \ref{fig:sim_manual} and \ref{fig:sim_control} show the results for the second scenario when the gas contents are chosen randomly from the combined dataset, instead of sequentially from separate datasets for each gas. In the first case, the power output is fairly stable, but the tank pressure is too high and the flares are activated more often. Additionally, the methane number limits are exceeded several times as the manually set mix ratio does not fall within the allowed interval. In the second case, the control inputs again oscillate more, but the duration of flaring is reduced by approximately 45\%. The methane number limits are slightly violated only once, when the gas contents are such that the methane number constraints cannot be met for any mix ratio. Also, there is no significant difference in the power output compared to the first case. Finally, the total cost of fuel is reduced by 19\% in the second case.

\balance

\section{Conclusion}
This paper presents artificial intelligence-based predictive fuel blending control for flare gas mitigation. Since the methane number is a nonlinear function that depends on gas composition, a neural network was trained to calculate it for different fuel blends and find the fuel mixing ratio values that satisfy the limits on methane number provided by the motor manufacturer. With these constraints on the mix ratio, a finite-horizon optimal control problem is solved where the main goal is to output the desired power while minimizing both the cost of the fuel and gas flaring for all possible control inputs. The optimal solution is then applied to the plant and the whole process is repeated. The algorithm was tested in two scenarios where the natural and permeate gas contents were sampled sequentially or randomly from the real data obtained from the site. The simulation results show that the proposed algorithm is capable of producing the desired power while mitigating costs and flaring, but also respecting the constraints on the fuel methane number.

\section*{Acknowledgment}
The authors would like to thank Bojan Spahija for his contribution to this work, especially in data analysis and plant modelling.

\bibliographystyle{IEEEtran}
\bibliography{bibliography}

\end{document}